\documentclass[conference]{IEEEtran}

\usepackage{orcidlink} 
\usepackage{cite}

\usepackage{graphicx} 

\usepackage{amssymb}

\PassOptionsToPackage{hyphens}{url}\usepackage{hyperref}
\usepackage{breakurl}

\usepackage[nameinlink, noabbrev, capitalize]{cleveref}
\usepackage{soul}
\usepackage{xcolor}

\begin{document}

\title{BoilerTAI: A Platform for Enhancing Instruction Using Generative AI in Educational Forums}
\date{May 2024}
\author{
    \IEEEauthorblockN{%
        Anvit Sinha\IEEEauthorrefmark{1},
        Shruti Goyal\IEEEauthorrefmark{1},
        Zachary Sy\IEEEauthorrefmark{1},
        Rhianna Kuperus\IEEEauthorrefmark{2}\orcidlink{0009-0002-3057-7671},
        Ethan Dickey\IEEEauthorrefmark{1}\orcidlink{0009-0007-3706-5253} and
        Andres Bejarano\IEEEauthorrefmark{1}\orcidlink{0000-0003-2611-2855}
    }%
    \IEEEauthorblockA{\IEEEauthorrefmark{1} \textit{Department of Computer Science}\\
        \textit{Purdue University}\\
        West Lafayette, IN 47906}
    \IEEEauthorblockA{\IEEEauthorrefmark{2} 
        \textit{Institutional Data Analytics and Assessment}\\
        \textit{Purdue University}\\
        West Lafayette, IN 47906}
}

\maketitle

\begin{abstract}

\textit{Contribution}: This Full paper in the Research Category track describes a practical, scalable platform that seamlessly integrates Generative AI (GenAI) with online educational forums, offering a novel approach to augment the instructional capabilities of staff. The platform empowers instructional staff to efficiently manage, refine, and approve responses by facilitating interaction between student posts and a Large Language Model (LLM). 

\textit{Background}: This study is anchored in Vygotsky's socio-cultural theory, with a particular focus on the concept of the More Knowledgeable Other (MKO). It examines how GenAI can augment the instructional capabilities of course staff in educational environments, acting as an auxiliary MKO to facilitate an enriched educational dialogue between students and instructors. This theoretical backdrop is important for understanding the integration of AI within educational contexts, suggesting a balanced collaboration between human expertise and artificial intelligence to enhance the learning and teaching experience.

\textit{Research Question}: How effective is GenAI in reducing the workload of instructional staff when used to pre-answer student questions posted on educational discussion forums?

\textit{Methodology}: Employing a mixed-methods approach, our study concentrated on select first and second-year computer programming courses with significant enrollments. The investigation involved the use of an AI-assisted platform by designated (human) Teaching Assistants (AI-TAs) to pre-answer student queries on educational forums. Our analysis includes a qualitative examination of feedback and interactions, focusing on the AI-TAs' experiences and perceptions. While we primarily analyzed efficiency indicators such as the frequency of modifications required to AI generated responses, we also explored broader qualitative aspects to understand the impact and reception of AI-generated responses within the educational context. This approach allowed us to gather insights into both the quantitative engagement with AI-assisted posts and the qualitative sentiments expressed by the instructional staff, laying the groundwork for further in-depth analysis.

\textit{Findings}: The findings indicate no significant difference in student reception to responses generated by AI-TAs compared to those provided by human instructors. This suggests that GenAI can effectively meet educational needs when adequately managed. Moreover, AI-TAs experienced a reduction in the cognitive load required for responding to queries, pointing to GenAI's potential to enhance instructional efficiency without compromising the quality of education.


\end{abstract}


\section{Introduction}

In the dynamic landscape of educational technology, adopting innovative tools has consistently opened new avenues for enhancing learning environments and instructional methodologies. As educational institutions face scalability and personalized learning challenges, Generative Artificial Intelligence (GenAI) has emerged as a transformative force, redefining educational possibilities and promising to enrich teaching and learning experiences \cite{lau2023BanIt, Becker2023programmingIsHard, Finnie22, Moradi23}. This paper introduces BoilerTAI, a novel platform integrating online educational forums with Generative Pre-trained Transformer (GPT) APIs. Through a centralized control panel, instructional staff can oversee and refine AI-generated responses to student posts before their release, ensuring the quality and relevance of interactions.

GenAI's ability to generate responsive content transforms instructional approaches in educational settings, particularly in managing large class sizes where individual student interaction is often limited. In such environments, students can use the course's online discussion forum to ask questions instead of attending crowded and time-constrained in-person office hours. GenAI could help handle the high volume of student posts, which typically surge before deadlines and exams, by providing first-draft responses to posts, reducing the cognitive load required of instructional staff to answer them. This study has found that GenAI can efficiently manage these inquiries, regardless of their frequency, and help ensure consistency and accuracy of responses. Furthermore, it can alleviate the tedium of answering repetitive and routine questions. In this way, GenAI allows Teaching Assistants (TAs) and Instructors to focus on more complex endeavors.


Regardless of the promising advantages of using GenAI in educational settings, its adoption introduces distinct challenges, primarily stemming from varying perceptions between students and instructors \cite{rogers24}. Instructors often adopt a cautious stance, highlighting potential risks related to accuracy, privacy, and the ethical implications of AI \cite{chan2023students, sheard24}. In contrast, students generally perceive GenAI technologies positively, valuing the personalized learning support and enhanced research capabilities they provide, though they also share concerns about AI's potential pitfalls \cite{amoozadeh24}.

Despite these mixed perceptions, the deployment of GenAI across educational domains has been swift and impactful \cite{olga2023generative, bull2023generative, archibald2023validation, chan2023students, liu24, Mai24}. Particularly relevant to this paper is using AI-based agents (i.e., chatbots) on discussion boards \cite{deakin2015ibm, goel2018jill, clarizia18, huang22, kai22, liu22, Subiyantoro23}. These agents notably reduce instructional workload by automating routine tasks such as grading and moderation of posts. Using chatbots streamlines administrative processes and allows educators to dedicate more attention to in-depth student interactions, thus helping address one of the fundamental challenges of modern education: maintaining high-quality teaching amidst ever-increasing class sizes \cite{archibald2023validation}.

In response to the challenges of scale and personalization in education, our proposed platform, BoilerTAI, uses the capabilities of OpenAI's GPT4.0 (GPT)\footnote{\url{https://openai.com/}} for educational discussion forums. BoilerTAI processes student queries by formulating prompts that guide GPT in generating pedagogically sound responses. These preliminary AI-generated answers are stored on a dashboard for an instructional staff's review instead of immediately posting them. This crucial step ensures that each response is vetted for accuracy, relevance, and educational value, allowing the instructional staff to request revisions through reprompting if necessary. Once an answer meets the staff's standards, it is approved and released on the forum, maintaining the appearance and authority of a direct response from that instructional staff.

This human-in-the-loop approach underpins BoilerTAI's operational philosophy, addressing critical concerns about the reliability of fully automated educational aids \cite{Jeon2023LargeLM}. BoilerTAI reduces the risk of disseminating incorrect or misleading information and preserves the educational experience's interpersonal element by incorporating human oversight, which is necessary for a successful learning experience \cite{ngo24, joshi24}. Students keep engaging with their discussion platform as if interacting with human instructors, which is important to prevent usage of this tool from disrupting regular student interaction. This method reflects a balanced integration of GenAI capabilities with human expertise, which is essential for enhancing educational outcomes while safeguarding the quality of instruction.

The theoretical underpinnings for BoilerTAI are rooted in Vygotsky's sociocultural theory, which emphasizes the fundamental role of social interaction in the development of cognition \cite{Vygotsky1978}. Central to this theory is the concept of the More Knowledgeable Other (MKO), traditionally a person with a better understanding or a higher ability level than the learner, facilitating the learner's cognitive development through targeted social interaction. In contemporary educational environments, where digital platforms and tools are becoming increasingly prevalent, the role of the MKO can also be extended to intelligent systems \cite{Stojanov2023}. One application of such can be theoretically seen as a Mindtool, the idea of which was developed by Jonassen in 1996. As they put it, ``mindtools are computer applications that, when used by learners to represent what they know, necessarily engage them in critical thinking about the content they are studying'' (p.1) \cite{jonassen1996computers, jonassen1998computers}. In order to use GenAI as an MKO, BoilerTAI addresses the issue of potentially misleading answers from these tools \cite{joshi24} by requiring the responses to be approved by one of the official instructors of the course.

A primary objective of this research is to evaluate the effectiveness of BoilerTAI in reducing the workload of instructional staff and enhancing the quality of responses within educational forums. The study specifically focuses on computer programming courses known for their high enrollment and complex nature of the content, leading to a high volume of student inquiries. By automating the initial response generation, BoilerTAI allows TAs and instructors to focus on refining and personalizing the support rather than crafting answers from scratch. This study aims to assess the impact of this tool on instructional efficiency in large enrollment computer science courses, where the demand for TAs is especially intense due to the volume of work and the students' diverse backgrounds and knowledge levels. In particular, using BoilerTAI we seek to answer the following question: \textit{How effective is GenAI in reducing the workload of instructional staff when used to pre-answer student questions posted on educational discussion forums?}

Our study adopted a mixed-methods approach to rigorously evaluate the effectiveness of BoilerTAI in four first and second-year computer science courses, each with substantial enrollments (around 200 each). This methodology combined quantitative analysis of response metrics and TA perceptions, measured through Likert-scale surveys \cite{likert1932}, with qualitative insights derived from the TAs' written feedback on their experiences using BoilerTAI. Key quantitative metrics included the the frequency of reprompts and the change in quality of TA responses. Qualitatively, we explored the detailed interactions and feedback from TAs to understand the more profound impacts of AI-assisted responses in the educational context. The preliminary findings are promising: BoilerTAI significantly enhanced the efficiency and quality of responses to student inquiries. Furthermore, the positive reception of AI-generated responses from students suggests that such technological integration can effectively augment educational interactions without compromising the learning experience.


\section{BoilerTAI Specifications}


\subsection{The BoilerTAI Platform}
BoilerTAI represents a significant advancement in how TAs interact with students over online discussion forums. BoilerTAI is a web-based application that integrates with Ed-Discussion \footnote{https://edtsem.org} to generate responses to student posts and inquiries using OpenAI's API in order to assist AI-TAs (TAs that utilize the BoilerTAI platform). The AI-generated responses are reviewed by the AI-TAs for approval, modification, or reprompting, when necessary, to ensure accuracy and appropriateness. 

While BoilerTAI functions as an intermediary platform, it is designed to minimize disruption to a TA's workflow. The approval workflow introduced into the AI's response generation process enhances the relevance of responses, ensures compliance with academic policies, and maintains a consistent tone with that of a TA. This AI-TA intervention is key to how BoilerTAI is meant to run.

BoilerTAI underwent a comprehensive five-month design, prototyping, and testing phase. This process involved collaboration with instructional specialists and computer science professors to ensure that the platform's prompt building and data collection features effectively utilized the strengths of OpenAI's API, focusing on usability and relevance.

\subsection{Technical stack}
The frontend is a static web application deployed on Surge.sh with Firebase Hosting, managed by a Cloudflare worker that handles load balancing through routing and caching strategies. The backend is a Dockerized Flask service deployed on Fly.io, which provides containerized micro-VMs across multiple regions globally\footnote{https://fly.io/}.

Firebase’s Cloud Firestore is used as the database, managing user states, authentication, synchronization between online discussion forums and AI, and data metrics collection. This NoSQL database efficiently handles data exporting and filtering in JSON format, ensuring scalability and real-time interaction capabilities.

OpenAI’s Python API is leveraged to build a pipeline that processes student questions from the discussion forum, retrieves necessary files, and integrates these elements into prompts. The responses are stored in the database, formatted for the user interface, and presented on the frontend.

Next.js is used to build a Tailwind-styled responsive user interface. Users can pull the latest student posts from Ed Discussion through their API, view posts with existing AI-generated responses, and approve responses for posting. They can manually edit AI-generated responses or select options for re-prompting.

The next layer utilizes Firebase’s Cloud Firestore as the database that manages user states and authentication, synchronization between online discussion forums and AI, and data metrics collection. As a NoSQL database, it seamlessly handles exporting and filtering data in JSON format, ensuring efficient scalability and real-time interaction capabilities.

Adjacent to the database, we leverage OpenAI’s Python API to build a pipeline. This system processes student questions from an online discussion forum, carries out necessary file retrieval tasks, and integrates these elements into our prompt. The received response is then stored in the database, formatted to fit the user interface, and presented on the frontend.

On the topmost layer, Next.js is used to build out a Tailwind-styled responsive user interface. Here, a user is able to pull the latest student posts from Ed Discussion through their API, view posts with existing AI-generated responses, and approve a response for posting. By viewing, they are able to manually edit the AI-generated response or select from options for re-prompting (discussed in further detail below).

\subsection{The BoilerTAI Workflow}
\begin{figure}[!t]
    \centering
    \includegraphics[width=8cm]{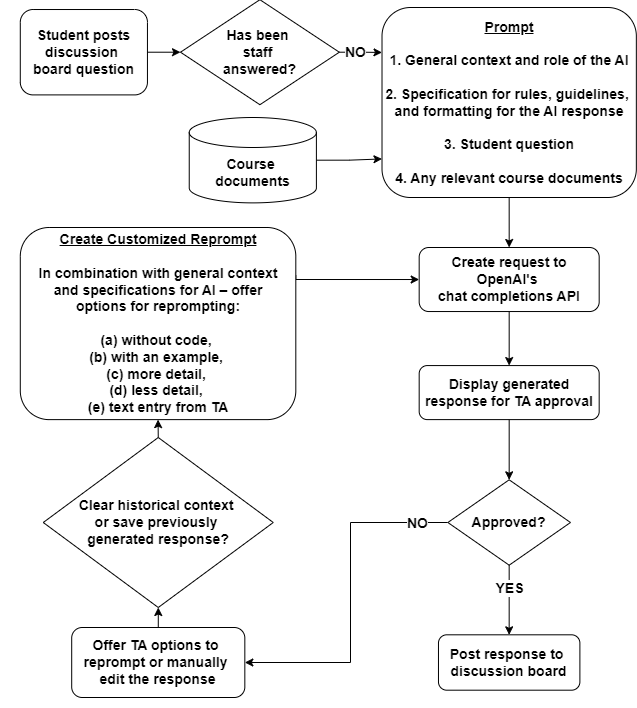}
    \caption{The generation of a response in BoilerTAI: The process begins with a student posting a question. The AI generates a response based on prompts and course documents, which is then reviewed and approved by AI-TAs before being posted. There are options for customized reprompting and manual editing to ensure accuracy and relevance.}
    \label{fig_boilertai_pipeline}
\end{figure}

Please refer to \Cref{fig_boilertai_pipeline} during the following discussion for the operational pipeline of the BoilerTAI platform. 

The operational workflow of the BoilerTAI platform begins when a student posts a question on the EdStem discussion board. Once the question is picked up on the automated periodic checks for unanswered queries, it is fed into the GPT-4 model. This step involves using a carefully crafted prompt, which is crucial for guiding the AI to generate contextually appropriate and pedagogically sound responses.

Once a response is generated, it is displayed on the BoilerTAI platform for an AI-TA to review. It is at this stage that our human-in-the-loop approach becomes pivotal. The AI-TAs play a critical role in ensuring the quality and appropriateness of the AI-generated responses. They have the authority to approve the response directly, reprompt the GenAI model in various ways, or choose to modify it directly all to better meet educational standards and the specific needs of the student.

The AI-TAs have several tools at their disposal to refine the AI's output via reprompting. They can decide to preserve the historical context, which involves using information from previously generated responses to inform current and future responses. Additionally, they can alter the response by allowing or disallowing code, adjusting the level of detail, or even crafting a completely custom prompt. This last option allows AI-TAs to provide specific context to the AI, guiding it to produce responses that are not only accurate but also aligned with the instructional goals of the course.

\subsection{Prompt Building}
Prompt engineering has been identified as a crucial element for optimizing the performance of AI tools like GPT, ensuring the generation of high-quality, contextually appropriate responses \cite{lo2023art}. The crafting of effective prompts is guided by principles outlined in the CLEAR Framework, a set of guidelines that underscores the significance of precision and clarity in communication, especially crucial in the Information Age \cite{lo2023clear}.

For BoilerTAI, using the CLEAR principles meant constructing a prompt that was straightforward and simple, focusing on facilitating problem-solving capabilities within students to enhance their self-sufficiency. This approach aligns with the broader educational objectives of ensuring independent learning and critical thinking among students.

To achieve these objectives, the prompt designed for BoilerTAI strictly defines the role of the AI as an objective teaching assistant in an undergraduate Computer Science course. The AI is configured to interact on discussion boards with a focus on guiding students towards solutions rather than providing direct answers. This ensures that the AI supports the learning process without bypassing the essential educational steps that students need to undertake themselves.

\section{Methods}
BoilerTAI was introduced as a controlled experiment in four first- and second-year computer science courses, each with a substantial enrollment of approximately 200 students. The primary objective was to assess the impact of BoilerTAI on the efficiency of Teaching Assistants (TAs) when used to answer student discussion board questions. In selecting first- and second-year computer science courses with high enrollments, the research specifically targeted settings where the volume and complexity of student inquiries could significantly benefit from AI intervention. These courses typically experience a high demand for teaching assistant resources and represent a critical testing ground for assessing the scalability and effectiveness of AI-assisted educational technologies. 

One TA per course, designated as the AI-TA, was tasked with exclusively using BoilerTAI to respond to discussion board queries. Notably, the remaining TAs and students were uninformed of BoilerTAI's existence and use, ensuring a controlled experimental environment through genuine interactions with the responses, as if they were written by the AI-TA themselves.

The study employed a mixed-methods approach to comprehensively analyze the impact of BoilerTAI. This approach was selected for its robustness in addressing complex research questions, allowing for a more nuanced understanding of both the statistical and thematic dimensions of GenAI's integration into educational settings. Quantitative analysis involved the examination of the AI-TAs' perceptions and experiences through Likert-scale surveys, as well as response metrics collected through the tool, including reasons for TA intervention (reprompting, editing, etc.). The full set of metrics are included in the full data analysis report, which will be made publicly available on arXiv.org upon publication. To maintain the authenticity of student interactions and avoid any potential bias due to knowing that a certain response may be AI generated, we chose not to inform nor survey students, ensuring that their engagement with the AI-generated responses was uninfluenced.

Furthermore, qualitative analysis explored the written responses provided by the AI-TAs, examining their feedback and interactions to better understand the impact and reception of AI-generated responses within the educational context.

The study design facilitated a preliminary evaluation of BoilerTAI's implementation and efficacy, offering valuable early insights into the integration of AI technologies in educational settings. Data collection was conducted during the Spring semester of 2024, coinciding with the academic schedule of the courses involved to ensure that the findings reflected the typical instructional conditions and student interactions during a standard academic period. This timing also allowed the research team to analyze the effects of the AI system under normal operational conditions, providing a realistic view of its potential and limitations in an actual educational setting.

Data on usage and refinement reasons were collected throughout the 3.5 month long trial. Two sets of surveys were conducted: an initial one to gauge the tool's performance, which was not formally analyzed, and a survey upon completion of the project, the analysis of which is presented in the following section.

\section{Results}
Following the piloting phase of BoilerTAI, four teaching assistants provided feedback through both multiple-choice responses and open-ended comments. The objective was to comprehensively assess the impact of BoilerTAI on users' experiences and perceptions. By soliciting feedback encompassing positive experiences, encountered challenges, suggested improvements, and usability enhancements, BoilerTAI endeavors to refine its features and user experience to better align with user needs. This mixed-methods data exploration serves as a foundational step towards optimizing BoilerTAI's functionality and efficacy in supporting both educators and learners.

\subsection{Efficiency and Quality of Responses}

The results of the study demonstrate that the implementation of BoilerTAI led to a significant enhancement in the efficiency of responses to student inquiries. Notably, 75\% (3 out of 4) of participants reported experiencing improvements across various degrees of magnitude, as outlined in \textbf{\Cref{tab:Question 3}}. Moreover, all respondents (4 out of 4) unequivocally acknowledged a discernible improvement in the quality of responses provided, as detailed in \textbf{\Cref{tab:Question 4}}. These empirical observations substantiate the primary objective of the investigation, which aims to optimize response efficiency while concurrently enhancing the overall quality of educational interactions.

\begin{table}
    \centering
    \caption{Results of Question 3: \textit{Since starting to use BoilerTAI, the efficiency \\ of my responses to student inquiries has improved.} }
    \label{tab:Question 3}

    \begin{tabular}{lcc}
         &  Count& Percent\\
         Strongly Disagree&  0& 00.0\\
         Somewhat Disagree&  1& 25.0\\
         Neutral&  0& 00.0\\
         Somewhat Agree&  3& 75.0\\
         Strongly Agree&  0& 00.0\\
         Total&  4& 100.0\\
    \end{tabular}
\end{table}

\begin{table}
    \centering
    \caption{Results of Question 4: \textit{Since starting to use BoilerTAI, the quality of my responses to student inquiries has improved.} }
    \label{tab:Question 4}

    \begin{tabular}{lcc}
         &  Count& Percent\\
         Strongly Disagree&  0& 00.0\\
         Somewhat Disagree&  0& 00.0\\
         Neutral&  0& 00.0\\
         Somewhat Agree&  4& 100.0\\
         Strongly Agree&  0& 00.0\\
         Total&  4& 100.0\\
    \end{tabular}
\end{table}

\subsection{Student Reception}
The reception of AI-assisted responses from students has been notably favorable. Analysis indicates that approximately 75\% (3 out of 4) of teaching assistants observed that responses generated through BoilerTAI elicited positive feedback from students (\textbf{\Cref{tab:Question 5}}). This finding underscores the effectiveness of AI integration in augmenting educational interactions and suggests that the utilization of AI technologies does not compromise the quality of the educational experience. Such considerations are critical for the widespread acceptance and implementation of such innovations within learning environments.

\begin{table}
    \centering
    \caption{Results of Question 5: \textit{In general, the responses from students to questions answered with the help of BoilerTAI have been positive.} }
    \label{tab:Question 5}

    \begin{tabular}{lcc}
         &  Count& Percent\\
         Strongly Disagree&  0& 00.0\\
         Somewhat Disagree&  0& 00.0\\
         Neutral&  1& 25.0\\
         Somewhat Agree&  3& 75.0\\
         Strongly Agree&  0& 00.0\\
         Total&  4& 100.0\\
    \end{tabular}
\end{table}

\subsection{Relevance and Personalization of Responses}
An overwhelming majority of the teaching assistants perceived the AI-generated responses as relevant and personalized. Specifically, approximately 75\% (3 out of 4) of respondents concurred that BoilerTAI’s responses were appropriately tailored to individual student inquiries (\textbf{\Cref{tab:Question 6}}). This observation underscores the significance of personalized feedback within educational settings, where tailored responses are pivotal in enhancing student understanding and engagement. Such findings highlight the importance of leveraging AI technologies to deliver tailored educational experiences that cater to the diverse needs of students.

\begin{table}
    \centering
    \caption{Results of Question 6: \textit{\textit{The BoilerTAI-generated responses to student inquiries are relevant and personalized.}} }
    \label{tab:Question 6}

    \begin{tabular}{lcc}
         &  Count& Percent\\
         Strongly Disagree&  0& 00.0\\
         Somewhat Disagree&  0& 00.0\\
         Neutral&  1& 25.0\\
         Somewhat Agree&  3& 75.0\\
         Strongly Agree&  0& 00.0\\
         Total&  4& 100.0\\
    \end{tabular}
\end{table}

\subsection{Need for Oversight}

However, it was also noted that there persists a need for oversight by the teaching assistants. Approximately 50\% (2 out of 4) of the teaching assistants reported requiring significant alterations or corrections to the responses suggested by BoilerTAI in approximately half of the instances (\textbf{\Cref{tab:Question 7}}). This observation highlights the crucial role of preserving a human element in the AI-assisted educational process. It underscores the necessity that responses are not solely factually accurate but also contextually appropriate. Such supervision ensures that educational interactions uphold the requisite depth and nuance essential for fostering effective learning outcomes.

\begin{table}
    \centering
    \caption{Results of Question 7: \textit{\textit{\textit{I find myself needing to correct or significantly alter the responses suggested by BoilerTAI:}}} }
    \label{tab:Question 7}
    
    \begin{tabular}{lcc}
         &  Count& Percent\\
         Never&  0& 00.0\\
         Sometimes&  1& 25.0\\
         About Half the Time&  2& 50.0\\
         Most of the Time&  1& 25.0\\
         Always&  0& 00.0\\
         Total&  4& 100.0\\
    \end{tabular}
\end{table}

\subsection{Potential for Further Integration}

The teaching assistants recognized a robust potential for further integrating BoilerTAI into educational practices. Seventy-five percent (3 out of 4) somewhat agreed, while 25\% (1 out of 4) strongly agreed that there exists significant potential for enhancing learning experiences through deeper integration of this AI tool into regular teaching practices (\textbf{\Cref{tab:Question 11}}). This sentiment not only reflects an optimistic outlook for the future of AI in education but also indicates an acknowledgment of its broader applications and potential for deeper integration across various educational scenarios. Such recognition underscores the importance of continued exploration and implementation of AI technologies to enrich educational experiences and improve learning outcomes.

\begin{table}
    \centering
    \caption{Results of Question 11: \textit{There is potential for further integrating BoilerTAI into educational practices to enhance learning experiences.} }
    \label{tab:Question 11}
    
    \begin{tabular}{lcc}
         &  Count& Percent\\
         Strongly Disagree&  0& 00.0\\
         Somewhat Disagree&  0& 00.0\\
         Neutral&  0& 00.0\\
         Somewhat Agree&  3& 75.0\\
         Strongly Agree&  1& 25.0\\
         Total&  4& 100.0\\
    \end{tabular}
\end{table}

\subsection{Open-Ended Responses}

The open-ended responses highlight how BoilerTAI enhances efficiency, reduces cognitive load, saves time, and effectively addresses certain types of student queries, particularly those related to course concepts. Users find that BoilerTAI simplifies the process of responding to student queries by reducing the cognitive load. Instead of needing to consider the formulation of an answer, its clarity, and correctness simultaneously, they can focus primarily on ensuring the accuracy of the response. This streamlining of the process makes it more efficient and less mentally taxing.

In particular, one participant noted:

\begin{quote}
    \textit{Prior to using the tool, I had to think about the answer to the question, phrasing that answer in a way that the student understands, and making sure that the final response is correct. Now, I mainly have to focus on only one of those three (ensuring correctness).}
\end{quote}

Another participant highlighted the time-saving aspect of BoilerTAI:

\begin{quote}
    \textit{It decreases the time it takes me to google a concept, type out a response and send it. The scaffold it creates for an answer is really helpful.}
\end{quote}

A third participant found BoilerTAI particularly helpful for addressing generic questions:
\begin{quote}
    \textit{It has been helpful in answering generic questions that are well described by students in the post related to concepts of the course.}
\end{quote}

\section{Discussion}

This study addresses the integration of Generative AI (GenAI) as an auxiliary More Knowledgeable Other (MKO)\cite{vygotsky1978mind}, drawing on Vygotsky’s socio-cultural theory to enhance interactions between students and instructors. Such integration enables TAs to focus more on complex educational dialogues requiring deeper cognitive involvement, while GenAI manages more routine interactions.

The following discussions stemming from our results demonstrate that BoilerTAI significantly enhances both the efficiency and quality of responses provided by AI-TAs. 

\subsection{Efficiency and Quality}
BoilerTAI has shown to significantly enhance the efficiency and quality of TA responses. According to the study, 75\% of AI-TAs reported improved efficiency in handling queries, while all participants noted an enhancement in response quality. Furthermore, the platform's capacity to generate scaffolds for answers has been particularly praised, as it aids in streamlining communication and reduces the cognitive load involved in crafting clear and accurate responses, enabling quicker and more organized exchanges between students and instructors. 

AI tools, while guiding students beyond their current understanding, typically lack the nuanced interactions human MKOs offer, such as using questioning techniques or providing partial solutions to promote deeper learning \cite{philbin2023exploring, de2023exploring, yadav2020assessment}. BoilerTAI mitigates this through advanced prompt engineering to ensure AI responses closely mimic those of human TAs, fostering engagement and critical thinking.

\subsection{Human Oversight and Ethical Considerations}
Positioning AI-TAs as intermediaries between students and the AI ensures essential human oversight, enhancing rather than replacing the human element in educational interactions. This strategy not only reduces the workload of TAs but also maintains the educational benefits of the MKO model. Despite the positive reception, with 75\% of AI-TAs noting student satisfaction, 50\% of AI-TAs still needed to correct or significantly alter AI responses about half the time. This underscores the importance of combining AI capabilities with human judgment to ensure the accuracy and contextual appropriateness of responses. Such an approach is also important for mitigating ethical concerns, particularly with maintaining academic integrity and ensuring that AI-generated responses do not mislead students or provide incorrect information \cite{smolansky2023educator}. However, it is important to keep in mind that as the field of GenAI grows, GenAI agents may be able to meet new ethical and accuracy standards (possibly even above TAs). We maintain that at the time of publication, GenAI agents are untested against such standards (in fact, one set of such standards has not been accepted by the general community) and therefore still require human oversight.

\subsection{Addressing Challenges and Limitations}
The BoilerTAI platform has been engineered to address specific challenges and limitations of utilizing AI in education, particularly in complex subject areas like science education. Issues such as the lack of context and evidence in AI-generated responses, concerns about academic integrity, potential for hallucinations (incorrect or misleading results) \cite{cooper2023examining, smolansky2023educator, baidoo2023education}, and concerns like the \textit{Junior-Year Wall} \cite{dickey2023innovating} are addressed by having TAs add context and corroborate facts, effectively saving time and ensuring reliability.

However, the study faced certain limitations, including a smaller TA sample size and the fact that not all TAs frequently answered questions on their discussion forums (greater than 2 days out of the week). These factors may have affected the generalizability of the findings and the overall evaluation of BoilerTAI's effectiveness. Nonetheless, the results should still be considered impactful, as there were consistent improvements in response efficiency and quality reported by the participating AI-TAs across their 4 distinct programming courses, which provides a strong indication of the platform's benefits.

\subsection{Future Directions}
The favorable reception of AI-assisted responses underscores BoilerTAI's effectiveness in maintaining high-quality educational interactions. However, the recent rise of GenAI leaves open many other research avenues. Future research could explore: student perceptions on our (educators') efforts to include (and restrict) GenAI in their education; the impact of our approaches on students' achievement, in particular if it allows us to add content to the curriculum or effectively ``raise the bar'' on expected student performance; similarly, if including and allowing GenAI in education allows us to move our learning objective for particular assessment types ``up'' on scales like the revised Bloom's Taxonomy \cite{revisedBloomsTaxonomy}; importantly, how do the educational principles that underpin tools like BoilerTAI translate to the K-12 space; and lastly on this non-comprehensive list (but perhaps the most interesting to the authors), how students' learning is impacted by (a) usage of AI and (b) our interventions that involve GenAI (including measures of learning from performance-based and cognitive/behavioral measures to affective and collaborative learning measures).  For BoilerTAI in particular, further research is essential to fully understanding the scalability of the human-in-the-loop approach and to refine AI's role in both enhancing efficiency and being a supportive tool within educational environments.

\section*{Acknowledgments}
This work was funded by Purdue's Innovation Hub (IH-AI-23002). The authors also acknowledge the Teaching Assistants (AI-TAs) from courses CS 180, CS 211, CS 251, and CS 253, who participated in the pilot program for BoilerTAI during the Spring of 2024.

\bibliographystyle{IEEEtran}
\bibliography{refs}

    
\end{document}